\documentclass[showpacs,preprintnumbers,amsmath,amssymb]{revtex4}

\usepackage{graphicx}
\usepackage{dcolumn}
\usepackage{bm}

\begin{document}

\title{In-medium properties of kaons in a chiral approach}

\author{Yue-Lei Cui and Bao-Xi Sun}

\affiliation{Institute of Theoretical Physics, College of Applied
Sciences, Beijing University of Technology, Beijing 100022, China }

\begin{abstract}
The first order self-energy corrections of the kaon in the symmetric
nuclear matter are calculated from kaon-nucleon scattering matrix
elements using a chiral Lagrangian within the framework of
relativistic mean field approximation. It shows that the effective
mass and the potential of  $K^+$ meson are identical with those of
$K^-$ meson in the nuclear matter, respectively.  The effective mass
of the kaon in the nuclear matter decreases with the nuclear density
increasing, and is not relevant to the kaon-nucleon Sigma term. The
kaon-nucleus potential is positive and increases with the nuclear
density. Moreover, the influence of the resonance $\Lambda(1405)$ on
the $K^-$-nucleus potential due to the re-scattering term is
discussed. Our results indicate the $K^-$ meson could not be bound
in the nuclei even if the contribution of $\Lambda(1405)$ resonance
is considered.
\end{abstract}
\pacs{12.39.Fe, 
      13.75.Jz, 
      21.65.+f. 
      }
\maketitle

The in-medium properties of kaons have caused more attentions of
nuclear physicists for many years. It was predicted 20 years ago
that Bose-Einstein condensation of $K^-$ mesons is possible in the
nuclear matter with a density up to several times of the normal
nuclear matter density, which is called kaon-condensation phenomenon
\cite{1}.  This prediction had inspired great interests of nuclear
physicists in the following years\cite{2,3,4,5}.  It's assumed that
there exists kaon-condensation in the high density nuclear matter in
the core of  neutron stars, and hence it can be utilized to soften
the equation of state of beta-stable matter in neutron stars
\cite{4,5}.

In 1999, by fitting the $K^-$ atomic data, it was predicted that
there can be ${\sl deeply~bound~kaonic~atoms}$, i.e., kaonic nuclei
\cite{6,7}. Although the experimental data at KEK and of FINUDA
collaboration were interpreted as the evidences that deeply bound
kaonic atoms exist\cite{8,8a,9,9a}, whether there can be deeply
bound states of $K^-$ meson in the nuclei is still an issue of great
controversy\cite{10,11,11a,11b}.

Whether or not there exist deeply bound states of $ K^- $ depends
upon properties of $K^- $ in the nuclear medium, and is closely
relevant to the depth of potential-well of $ K^- $ in the nuclear
matter. Nevertheless, it can be seen from present results that the
interaction between the kaon and the nucleus is mainly
model-dependent. By fitting the data of $K^-$ atoms, a strong
attractive $K^-$-nucleus potential with the depth of 150-200MeV is
obtained\cite{12}. However, the calculation based on the chiral
coupling channel approach gives the $K^-$-nucleus potential ranging
from 85-140MeV\cite{12a,13,13a}, and the chiral unitary theory that
starts from the bare $K^-N$ interaction predicts an even more
shallower $K^-$-nucleus potential in the range of 50 -70
MeV\cite{14,15}. The medium properties of kaons have also been
studied from mean field theories, built within the framework of
chiral Lagrangians\cite{16,17}, based on Walecka model extended to
incorporate kaons\cite{18,19}, or using explicitly quark degrees of
freedom\cite{20}.
Moreover, the behavior of the effective masses of kaons at the
finite nuclear density has been studied in the $SU(3)$
Nambu-Jona-Lasinio model\cite{Co02}.

Although these models give different $K^-$-nucleus  potential
depths, most of them predict negative values of the $K^-$-nucleus
potential in the nuclear matter, i.e., the $ K^- $ meson will $feel$
an attractive force in the nucleus. Moreover, many models predict
different behaviors of $ K^+ $ and $ K^- $ mesons in the nuclear
matter. The effective energy of $ K^+ $ meson is higher than its
energy in vacuum, and increases with the nuclear density.
Contrarily, the effective energy of $ K^- $ meson is lower than its
energy in vacuum, and decreases with the nuclear density
\cite{18,19,21}. Therefore, it is easy to draw a conclusion that $
K^+ $ meson can not be bound in the nucleus and whether $ K^- $
meson can be bound in the nucleus or not will be determined by the
correct calculation on the $K^-$-nucleus potential. In Ref.
~\cite{13}, T. Waas and W. Weise give an increasing effective mass
of $ K^+ $ meson with the nuclear density. However, some other
models give different results.

In our previous papers, according to Wick's theorem, we have studied
the in-medium properties of the photon, the scalar meson and the
vector meson\cite{22}. In this paper, we will calculate the
self-energy of the kaon in the nuclear matter from kaon-nucleon
($KN$) scattering matrix elements using a chiral Lagrangian, and
then study the in-medium properties of kaons in the nuclear matter.

According to the one-boson exchange theory, nucleons interact with
each other by exchanging mesons. The Walecka model provides
nucleon-meson coupling Lagrangian density in the
form\cite{Wa74,Re89,Ri96,Me06}:
\begin{eqnarray}
\label{eq:Lagr}
 {\cal L}~&=&~\bar\psi\left(i\gamma_{\mu}\partial^{\mu}
 -M_N\right)\psi-g_\sigma\bar\psi\sigma\psi-g_\omega\bar\psi \gamma_\mu
\omega^\mu\psi    \nonumber \\
 &&~+~\frac{1}{2}\partial_\mu\sigma\partial^\mu\sigma-U(\sigma)
 -\frac{1}{4}\omega_{\mu\nu}\omega^{\mu\nu}
+ \frac{1}{2}m^2_\omega\omega_\mu\omega^\mu
\end{eqnarray}
with
\begin{equation}
U(\sigma)=\frac{1}{2} m^2_\sigma
\sigma^2_{}+\frac{1}{3}b\sigma^3_{}+\frac{1}{4}c\sigma^4_{},
\end{equation}
where $\psi$, $\sigma$ and $\omega$ denote field operators of the
nucleon, the scalar meson and the vector meson, respectively. The
contribution of iso-vector mesons to the equation of state of the
symmetric nuclear matter is zero within the framework of
relativistic mean-field (RMF) approximation, so we only considered
the scalar meson $\sigma$ and the vector meson $\omega$ in
Eq.~(\ref{eq:Lagr}).

In the relativistic mean-field approximation, the effective mass and
effective energy of the nucleon are defined as:
\begin{eqnarray}
M_N^*=M_N+g_\sigma\sigma_0, \nonumber
\end{eqnarray}
and
\begin{equation}
\varepsilon^{(+)}(\vec{p})=\sqrt{\vec{p}^2+{M_N^*}^2}+g_\omega\omega_0,
\nonumber
\end{equation}
where $\sigma_0$, $\omega_0$ are the expectation values of the
scalar meson $\sigma$ and the vector meson $\omega$ in the nuclear
matter, respectively.

In the chiral expansion for mesons and baryons, the $KN$ interaction
Lagrangian to the next-to-leading order can be written as
\cite{23,24,25}:
\begin{equation}
\label{eq:knlag} {\cal
L}^{chiral}_{KN}~=~g_1\bar{\psi}\gamma^\mu\psi[(\partial_{\mu}\bar{K})K
-\bar{K}\partial_{\mu}K]
~+~g_2\bar{\psi}\psi\bar{K}K
~+~g_3\bar{\psi}\psi(\partial_{\mu}\bar{K}\partial^{\mu}K),
\end{equation}
where
\begin{equation}
K=\begin{pmatrix} K^+ \\  K^0  \end{pmatrix},\hspace{60pt} \bar
K=\begin{pmatrix} K^-, & \bar{K^0}  \end{pmatrix}, \nonumber
\end{equation}
and the coupling constants are:
\begin{equation}
\label{couple const}
    g_1=\frac{3i}{8f^2_K},~~~~g_2=\frac{\Sigma_{KN}}{f^2_K},
    ~~~~g_3\approx(\frac{0.33}{m_K}-\frac{\Sigma_{KN}}{m^2_K})/{f^2_K}
     \nonumber
\end{equation}
with $f_K\approx93MeV$ the decay constant of the kaon and
$\Sigma_{KN}$ the Sigma term of $KN$ interaction. Because we are
studying the properties of the kaon in the symmetric nuclear matter,
the nucleon isospin correlating terms have been ignored in the $KN$
interaction Lagrangian.  In Eq.~(\ref{eq:knlag}), the first term
corresponds to Tomozawa-Weinberg vector interaction, the second term
is $KN$ scalar interaction term and the last is the off-shell term.

The nucleon field operator and its conjugate operator can be
expanded in terms of a complete set of solutions to the Dirac
equation:
\begin{eqnarray} \label{eq:psi}
\psi(x)&=&\sum_{\lambda=1,2}\int\frac{d^{3}p}{(2\pi)^{\frac{3}{2}}}
\sqrt{\frac{M^*_N}{E^*(\vec{p})}}
[A_{\vec{p}\lambda}U(\vec{p},\lambda)
\exp\left(i\vec{p}\cdot\vec{x}-i\varepsilon^{(+)}(\vec{p})t\right),
\end{eqnarray}
and
\begin{eqnarray} \label{eq:bpsi}
\bar{\psi}(x)&=&\sum_{\lambda=1,2}\int\frac{d^{3}p}{(2\pi)^{\frac{3}{2}}}
\sqrt{\frac{M^*_N}{E^*(\vec{p})}}
[A^\dagger_{\vec{p}\lambda}\bar{U}(\vec{p},\lambda)
\exp\left(-i\vec{p}\cdot\vec{x}+i\varepsilon^{(+)}(\vec{p})t\right),
\end{eqnarray}
where $E^*(\vec{p})=\sqrt{\vec{p}^2+{M^*_N}^2}$, and $\lambda$
denotes the spin of the nucleon, $A_{\vec{p}\lambda}$ and
$A^\dagger_{\vec{p}\lambda}$ are the annihilation and creation
operators of the nucleon, respectively. We have assumed that there
are not antinucleons in the ground state of nuclear matter, thus
only positive-energy components are considered in Eqs.
~(\ref{eq:psi}) and ~(\ref{eq:bpsi}).

The kaon field operators with the fixed momentum $k$ can be
expressed as
\begin{eqnarray}
\label{eq:kfixed}
K^+(k,x)&=&a_{K^+}(k)e^{-ik\cdot x}+a^{\dag}_{K^-}(k)e^{ik\cdot x},\nonumber \\
K^-(k,x)&=&a_{K^-}(k)e^{-ik\cdot x}+a^{\dag}_{K^+}(k)e^{ik\cdot x},\nonumber\\
K^0(k,x)&=&a_{K^0}(k)e^{-ik\cdot x}+a^{\dag}_{\bar{K^0}}(k)e^{ik\cdot x},\nonumber\\
\bar{K}^0(k,x)&=&a_{\bar{K}^0}(k)e^{-ik\cdot
x}+a^{\dag}_{K^0}(k)e^{ik\cdot x}.
\end{eqnarray}
With the Legendre transformation of the Lagrangian in
Eq.~(\ref{eq:knlag}), the $KN$ interaction Hamiltonian can be
written as
\begin{equation}
{\cal H}_I(x)={\cal H}_1(x)+{\cal H}_2(x)+{\cal H}_3(x),
\end{equation}
where
\begin{eqnarray}
\label{eq:kfield}
{\cal H}_1&=&g_1\bar{\psi}\gamma^j\psi[\bar{K}\partial_{j}K-(\partial_{j}\bar{K})K],\nonumber\\
{\cal H}_2&=&-g_2\bar{\psi}\psi\bar{K}K,\nonumber\\
{\cal H}_3&=&g_3\bar{\psi}\psi(\dot{\bar{K}}\dot
K-\partial_{j}\bar{K}\partial^{j}K).
\end{eqnarray}
In order to obtain the self-energies of kaons in the nuclear matter,
the first order $KN$ scattering matrix should be calculated firstly:
\begin{eqnarray}
\label{eq:smatrix}
{\hat{S}}_1&=&-i\int d^4x T[{\cal H}_I(x)]\nonumber\\
&=&-i\int d^4x T\{{\cal H}_1(x)+{\cal H}_2(x)+{\cal H}_3(x)\}.
\end{eqnarray}
The corresponding Feynman diagram is shown in Fig.~1. At the point
of the saturation density of the nuclear matter, the first order
self-energy correction is enough for us to study the in-medium
properties of kaons.

By substituting
Eqs.~(\ref{eq:psi}),~(\ref{eq:bpsi}),~(\ref{eq:kfixed}) and
~(\ref{eq:kfield}) into Eq.~(\ref{eq:smatrix}), the first order
scattering matrix element for $ K^- $ meson in the nuclear matter
can be written as
\begin{eqnarray}
\label{eq:ksm}
&& \langle~p\lambda,k~|~\hat{S}_1~|~p\lambda,k~\rangle  \nonumber\\
&=&-i(2\pi)^4 \{-2ig_1\int\frac{d^{3}p}{(2\pi)^3}\frac{4\vec{p}\cdot\vec{k}}
{2E^*(\vec{p})}\theta(p_F-|\vec{p}|)\nonumber\\
&&+[-g_2+g_3({\omega}^2_k+\vec{k}^2)]\int\frac{d^{3}p}{(2\pi)^3}\frac{2M_N^*}
{E^*(\vec{p})}\theta(p_F-|\vec{p}|)\}\nonumber\\
&=&-i(2\pi)^4\{0+[-g_2+g_3({\omega}^2_k+\vec{k}^2)]\rho_s\}.
\end{eqnarray}
Considering the number density of protons equals that of neutrons in
the symmetric nuclear matter, the scalar density of nucleons
$\rho_s$ in Eq.~(\ref{eq:ksm}) should be replaced with:
\begin{equation}
\rho_s=4\int\frac{d^{3}p}{(2\pi)^3}\frac{M_N^*}{\sqrt{\vec{p}^2+{M_N^*}^2}}
\theta(p_F-|\vec{p}|),
\nonumber
\end{equation}
where $p_F$ is the Fermi momentum of nucleons in the nuclear matter.
With the Dyson equation for the propagator of kaons in the nuclear
matter
\begin{equation}
\frac{i}{k^2-m_K^2+i\varepsilon-\Sigma_K}
=\frac{i}{k^2-m_K^2+i\varepsilon}+\frac{i}
{k^2-m_K^2+i\varepsilon}(-i\Sigma_K)\frac{i}{k^2-m_K^2+i\varepsilon},
\nonumber
\end{equation}
the first order self-energy correction of $K^-$ meson is obtained
as:
\begin{equation}
\label{eq:self}
-i\Sigma_K=-i[-g_2+g_3({\omega}^2_k+\vec{k}^2)]\rho_s.
\end{equation}
Evidently, the Tomozawa-Weinberg interaction term in
Eq.~(\ref{eq:knlag}) has no contribution to the first order
self-energy correction of $K^-$ meson in the nuclear matter. The
first order self-energy correction of $K^+$, $K^0$ and $\bar K^0$
mesons in the nuclear matter can be obtained similarly, and the
results are same as that of $K^-$  meson in Eq.~(\ref{eq:self}).

Taking into account the on-shell relation of the kaon
\begin{equation}
\label{eq:onshell}
\omega^2_k=\vec{k}^2+m_K^2, \nonumber\\
\end{equation}
the effective energy and effective mass of the kaon in the nuclear
matter can be written as
\begin{equation}
\omega^*=(1-2g_3\rho_s)^{1/2}\omega_k=[1-2\rho_s(\frac{0.33}{m_K}-\frac{\Sigma_{KN}}{m^2_K})
/{f^2_K}]^{1/2}\omega_k,
\end{equation}
and
\begin{equation}
\label{eq:kmasun}
m_K^*=[m_K^2-(m_K^2g_3+g_2)\rho_s]^{1/2}=[m_K^2-\frac{0.33m_K\rho_s}{f^2_K}]^{1/2},
\end{equation}
respectively.

In a chiral approach, we found that the effective energy and
effective mass of kaons and anti-kaons in the nuclear matter are
identical, respectively. Moreover, the effective mass of the kaon is
independent of the $KN$ Sigma term $\Sigma_{KN}$. Such result is
different from previous models.

The effective mass of the kaon $m_K$ as a function of nucleon
density $\rho$ in the framework of relativistic mean-field
approximation is shown in Fig.~2, where $\rho_0=0.15fm^{-3}$ is the
saturation density of the normal nuclear matter. The solid line
denote the results with NL3 parameters\cite{26}, and the dashed line
is for the NLSH parameters\cite{27}. The effective kaon mass
decreases monotonically with the nucleon density increasing. At the
point of the saturation density, the effective kaon mass is 473MeV,
about 20MeV less than the corresponding value in vacuum, The results
for the two different parameter sets are almost same as each other
when $\rho<1.5\rho_0$. At higher densities for $\rho>1.5\rho_0$, the
effective mass of the kaon is nearly a constant for NL3 parameters,
while the result with NLSH parameters decreases continually.

In Ref.~\cite{19}, the effective mass of the kaon in the nuclear
matter is extracted straightly from the equation of motion of the
kaon in the relativistic mean-field approximation:
\begin{equation}
\label{eq:kmasch} m_K^*=\sqrt{\frac{m_K^2-g_2\rho_s}{1+g_3\rho_s}}.
\end{equation}
At the low density limit $g_3\rho_s\ll1$,
\begin{eqnarray}
m_K^*&=&\sqrt{\frac{m_K^2-g_2\rho_s}{1+g_3\rho_s}}
=\sqrt{(m_K^2-g_2\rho_s)[1-g_3\rho_s+o(g_3\rho_s)]}\nonumber\\
&\approx &\sqrt{m_K^2-g_2\rho_s-m_K^2g_3\rho_s}. \nonumber
\end{eqnarray}
The equation of the effective mass of the kaon in
Eq.~(\ref{eq:kmasch}) is just in agreement with our result in
Eq.~(\ref{eq:kmasun}).

The kaon-nucleus potential is defined as the difference between the
effective energy of the kaon in the nuclear matter and its energy in
vacuum  at the limit of zero momentum,
\begin{eqnarray}
\label{eq:uk}
U_K&=&\omega^*(\vec{k})\mid_{\vec{k}=0}-\omega(\vec{k})\mid_{\vec{k}=0}
=(1-2g_3\rho_s)^{1/2}m_K-m_K  \nonumber \\
&=&[1-2\rho_s(\frac{0.33}{m_K}-\frac{\Sigma_{KN}}{m^2_K})/{f^2_K}]^{1/2}m_K-m_K.
\end{eqnarray}
The kaon-nucleus potential is relevant to the $KN$ Sigma term
$\Sigma_{KN}$. In the original paper\cite{25}, the authors choose
$\Sigma_{KN}\approx2m_\pi$ in accordance with the Bonn
model\cite{28}, while the value $\Sigma_{KN}=450\pm30MeV$ is favored
according to lattice gauge calculations\cite{29}. The kaon-nucleus
potentials $U_K$ at different nuclear matter densities are
illustrated in Fig.~3.

It can be seen clearly that both $K^+$ meson and $K^-$ meson have
positive potentials in the nuclear matter. Meanwhile, the
kaon-nucleus potential $U_K$ increases with the nucleon number
density $\rho$.
It means the effective energy of the kaon in the nuclear matter is
higher than the corresponding energy of the kaon in vacuum, and the
the effective energy of the kaon at the fixed momentum increases
with the nuclear density.
$U_K$ is relevant to the $KN$ Sigma term $\Sigma_{KN}$. At the
saturation density $\rho=\rho_0$, $U_K=44MeV$ for
$\Sigma_{KN}=450MeV$, and $U_K=6.5MeV$ for $\Sigma_{KN}=280MeV$. In
the range of $\rho<1.5\rho_0$, the kaon-nucleus potential $U_K$ is
nearly independent of the choice of different RMF parameter sets. As
$\rho>3\rho_0$, the increase of $U_K$ becomes slow. At
$\rho=3\rho_0$,  $U_K=70MeV$ for $\Sigma_{KN}=450MeV$ and $U_K=8MeV$
for $\Sigma_{KN}=280MeV$. The positive values of the kaon-nucleus
potential in Fig.~3 imply neither $K$ meson nor $\bar K$ meson can
be bound in the nuclei. Obviously, our results do not support the
occurrence of kaon condensation in the high density nuclear matter,
either.

If the self-consistency is taken into account in the calculation of
$U_K$,  the value of kaon mass $m_K$ in the effective energy of the
kaon $\omega^{*}(\vec{k})$ in Eq.~(\ref{eq:uk}) should be replaced
with the corresponding effective mass of the kaon $m_K^{*}$ in the
nuclear matter,  and then the kaon-nucleus potential in the nuclear
matter will take the form:
\begin{eqnarray}
\label{eq:ukeff}
U_K&=&(1-2g_3\rho_s)^{1/2}m_K^*-m_K  \nonumber \\
&=&[1-2\rho_s(\frac{0.33}{m_K}-\frac{\Sigma_{KN}}{m^2_K})/{f^2_K}]^{1/2}
[m_K^2-\frac{0.33m_K\rho_s}{f^2_K}]^{1/2}-m_K \nonumber \\
&\approx&\rho_s(\frac{\Sigma_{KN}}{m_K}-\frac{0.99}{2})/{f^2_K}.
\end{eqnarray}
From Eq.~(\ref{eq:ukeff}) we can see even if the $KN$ Sigma term
takes its minimum value $\Sigma_{KN}=280MeV$, the kaon-nucleus
potential is still positive, i.e.,
${U_K}_{min}=U_K(\Sigma_{KN}=280MeV)>0$. It shows once more that
there can not be kaonic bound states in the nucleus.

In this paper, we utilized the $KN$ interacting Lagrangian in a
chiral approach to figure out the first order self-energy correction
of the kaon in the nuclear matter. The calculation results show the
effective mass and the potential of $K$ meson are identical with
those of $\bar K$ meson in the nuclear matter, respectively.  The
effective mass of the kaon in the nuclear matter decreases with
nuclear density increasing, and is not relevant to the $KN$ Sigma
term.  The potential of the kaon in the nuclear matter is positive
and increases with the nuclear density. In a conclusion, as long as
the $KN$ interacting Lagrangian given in Eq.~(\ref{eq:knlag}) is
correct, it's impossible that there exist $K^-$ bound states in the
nuclei. Probably the kaon may react with nucleons to form a hyperon
in the nucleus. Meanwhile, neither do our results support the
possibility of kaon condensation in the denser nuclear matter.

In Ref.~\cite{10,14}, the influence of the resonance $\Lambda(1405)$
on the $K^-$-N interaction is studied and a $K^-$-nucleus potential
with a depth about 50MeV and a width about 100MeV is obtained in the
chiral unitary approach. In order to calculate the re-scattering
term including the intermediate bound state $\Lambda(1405)$ due to
the Weinberg-Tomozawa interaction in Eq.~(\ref{eq:knlag}), the
$\Lambda(1405)$ field operator $\Lambda$ and its Dirac conjugate
operator $\bar \Lambda$ are assumed to be
\begin{equation}
    \Lambda=\frac{\psi\bar K}{f_K},~~~~~\bar\Lambda=\frac{K \bar\psi}{f_K},
\end{equation}
respectively. Therefore, the Weinberg-Tomozawa term in
Eq.~(\ref{eq:knlag}) can be rewritten as
\begin{equation}
\label{eq:lambda1405} {\cal
L}_{K\Lambda}~=~g_1^\prime[(\partial_{\mu}\bar{K})\bar{\Lambda}\gamma^\mu\psi
-\bar{\psi}\gamma^\mu\Lambda\partial_{\mu}K]
\end{equation}
with $g_1^\prime={3i}/{8f_K}$.
The 2nd order self-energy correction of $K^-$ meson in the nuclear
matter due to Eq.~(\ref{eq:lambda1405}) is obtained via
\begin{equation}
\label{eq:klambdaself} \Sigma_{K\Lambda}~=~-{g_1^\prime}^2 \int
\frac{d^3p}{(2\pi)^3}\left[\frac{Tr[(\rlap{/}p+M_N)\gamma^ik_i(\rlap{/}k+\rlap{/}p+M_\Lambda)\gamma^jk_j]}
{2\sqrt{p^2+M_N^2}~[(k+p)^2-M^2_\Lambda]}~\theta(p_F-|\vec{p}|)
\right] \nonumber
\end{equation}
with $k_i$ the momentum of the $K^-$ meson and $M_\Lambda$ the mass
of the resonance $\Lambda(1405)$.
At the low momentum limit $|\vec{k}|/m_K\approx0$, the effective
energy and effective mass of $K^-$ meson in the nuclear matter can
be written as
\begin{equation}
\omega^\ast_k~=~\omega_k~f(\rho)
\end{equation}
and
\begin{equation}
m^\ast_K~=~m_K~f(\rho)
\end{equation}
with
\begin{equation}
f(\rho)~=~\sqrt{\left[1-\frac{{g_1^\prime}^2}{2\pi^2}\int_0^{p_F}p^2dp\frac{1}{\sqrt{p^2+M_N^2}}\left(
1-\frac{4p^2}{3(m_K^2+M_N^2-M_\Lambda^2+2\omega_k
\sqrt{p^2+M_N^2})}\right) \right]}.
\end{equation}
Therefore, the $K^-$-nucleus potential relevant to the intermediate
bound state $\Lambda(1405)$ is defined as
\begin{equation}
U_{K\Lambda}~=~\lim_{\omega_k\rightarrow m_K} m_K[f(\rho)~-~1].
\end{equation}
At the saturation density of the symmetric nuclear matter,
$\rho=\rho_0$, $U_{K\Lambda}\approx 0.57MeV$. If the mean-field
effect is taken into account in the calculation, the value of
$U_{K\Lambda}$is still positive and changes slightly.
In conclusion, our results imply the $K^-$ meson could not be bound
in the nuclei, even if the contribution of $\Lambda(1405)$ resonance
is considered.

The authors are grateful for useful discussions with Y. -C. Huang,
U. Lombardo, Z. -H. Xiong, E. -G. Zhao, P. -Z. Ning, L. Li, X. -H.
Zhong and X. -F. Lu. We would also like to acknowledge comments and
suggestions made by an anonymous referee. This work was supported in
part by the Foundations of Beijing University of Technology, the
Funding Project for Academic Human Resources Development in
Institutions of Higher Learning Under the Jurisdiction of Beijing
Municipality and the Scientific Research Foundation for the ROCS,
State Education Ministry.

\newpage

\begin{figure*}
\includegraphics{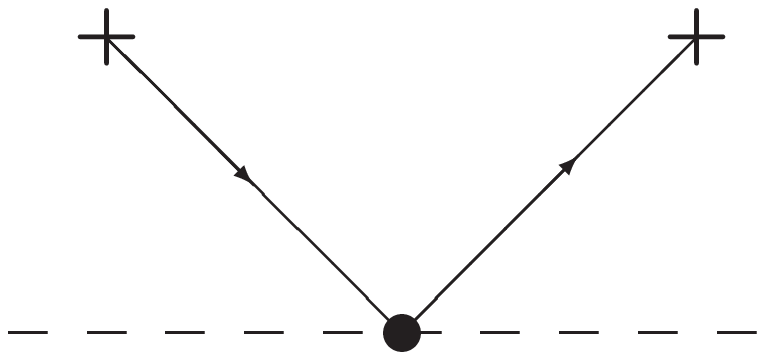}
\caption{\label{fig1}Feynman diagram on the first order self-energy
of the kaon in the nuclear matter. The dashed lines denote the kaon,
and the solid lines denote the nucleon.}
\end{figure*}

\begin{figure*}
\includegraphics{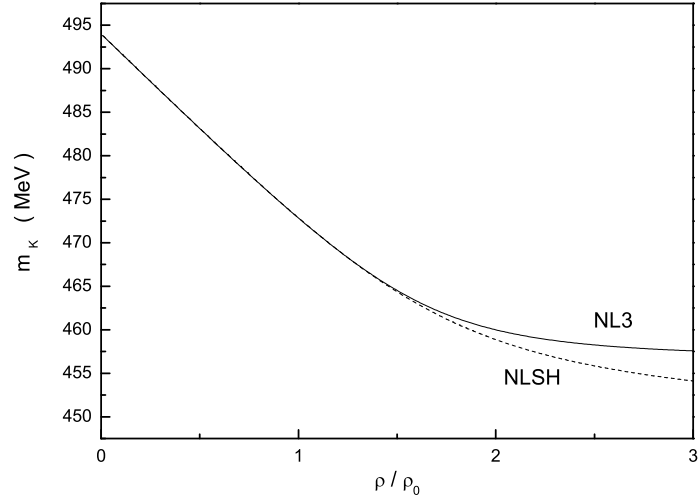}
\caption{\label{fig2}The effective mass of the kaon $m_K$ as a
function of nucleon density $\rho$ in units of saturation densities
$\rho_0$ of the nuclear matter for different parameter sets, and
$\rho_0=0.15fm^{-3}$. The solid line denotes the results with NL3
parameters, and the dashed line is for the NLSH parameters.}
\end{figure*}

\begin{figure*}
\includegraphics{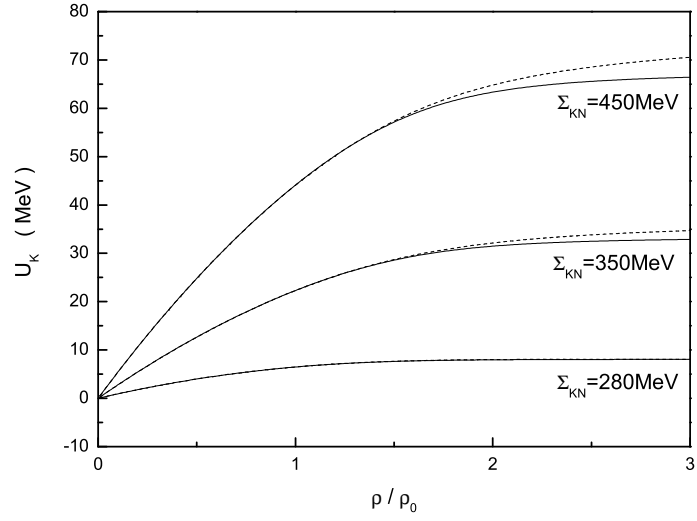}
\caption{\label{fig3} The potential of the kaon $U_K$ in the nuclear
matter as a function of nucleon density $\rho$ in units of
saturation densities $\rho_0$ of the nuclear matter for different
values of the $KN$ sigma term, and $\rho_0=0.15fm^{-3}$. The
meanings for the solid and dashed lines are same as those in Fig.
2.}
\end{figure*}

\end{document}